\documentclass[11pt]{article}
\usepackage{amsmath}
\usepackage{amssymb}
\usepackage{graphicx}
\usepackage{color}
\usepackage{cancel}
\usepackage{fancybox}

\numberwithin{equation}{section}

\begin{document}

\newcommand{\ba}{\begin{array}}
\newcommand{\ea}{\end{array}}

\renewcommand{\thefootnote}{\fnsymbol{footnote}}

%
\newcommand{\bd}{\begin{displaymath}}
\newcommand{\ed}{\end{displaymath}}
\newcommand{\bi}{\begin{itemize}}
\newcommand{\ei}{\end{itemize}}
\newcommand{\benu}{\begin{enumerate}}
\newcommand{\eenu}{\end{enumerate}}
\newcommand{\be}{\begin{equation}}
\newcommand{\ee}{\end{equation}}
\newcommand{\bea}{\begin{eqnarray}}
\newcommand{\eea}{\end{eqnarray}}
\def\lsim{\ \raisebox{-.45ex}{\rlap{$\sim$}} \raisebox{.45ex}{$<$}\ }
\def\gsim{\ \raisebox{-.45ex}{\rlap{$\sim$}} \raisebox{.45ex}{$>$}\ }
\newcommand{\eqs}{Eqs.}
\newcommand{\Def}{Definition}
\newcommand{\fig}{Figure}
\newcommand{\Fig}{Figure}
\newcommand{\figs}{Figures}
\newcommand{\Figs}{Figures}
\newcommand{\Ref}{Ref.~}
\newcommand{\Refs}{Refs.~}
\newcommand{\Sec}{Section}
\newcommand{\Secs}{Sections}
\newcommand{\App}{Appendix}
\newcommand{\Apps}{Appendices}
\newcommand{\tab}{Table}
\newcommand{\tabs}{Tables}
\newcommand{\Tab}{Table}
\newcommand{\Tabs}{Tables}
\newcommand{\equ}[1]{\eq~(\ref{equ:#1})}
\newcommand{\figu}[1]{\fig~\ref{fig:#1}}
\newcommand{\eV}{\mbox{ eV}}
\newcommand{\sss}{\sin^2 \theta_{12}}
\def\nn{\nonumber}
\newcommand{\etald}{{\it et~al}}
%
%
\def\nue{{\nu_e}}
\def\anue{{\bar\nu_e}}
\def\numu{{\nu_{\mu}}}
\def\anumu{{\bar\nu_{\mu}}}
\def\nutau{{\nu_{\tau}}}
\def\anutau{{\bar\nu_{\tau}}}
%
%
\def\gev{{\rm GeV~}}
\def\gevd{{\rm GeV}}
\def\evsq{{{\rm eV$^2$~}}}
\def\evsqd{{{\rm eV$^2$}}}
%
%
\newcommand{\dms}{\mbox{$\Delta m^2_{\odot}$}}
\newcommand{\dma}{\mbox{$\Delta m^2_{\rm 31}$}}
\def\dcp{{\delta_{\mathrm{CP}}}}
\def\stsmall{{\sin^2 2 \theta_{13}}}

\newcommand{\obb}{0\mbox{$\nu\beta\beta$}}
\newcommand{\meff}{\mbox{$\langle m \rangle$}
}
\begin{titlepage}
\begin{flushright}
HRI-P-08-07-004 \\
\end{flushright}

\begin{center}
\vspace*{11mm}
{\Large \bf Minimal Seesaw Textures with Two Heavy Neutrinos 
}
\vspace{.5in}

Srubabati Goswami$^{1,2}$\footnote{sruba@prl.res.in}, 
~Atsushi Watanabe$^1$\footnote{watanabe@mri.ernet.in}
\vskip 0.5cm
$^1${\small {\it Harish-Chandra Research Institute, Chhatnag Road, Jhunsi,
Allahabad  211 019, India}},\\
$^2${\small {\it Physical Research Laboratory, Navrangpura, Ahmedabad -380009, India}},\\

\vskip 1in

\end{center}

\begin{abstract}\noindent%
We systematically analyze the Dirac and the Majorana mass matrices
in seesaw models with two heavy right-handed neutrinos.
We perform thorough classification of the vanishing matrix elements
which are compatible with the results from the current neutrino
oscillation experiments.
We include the possibility of a non-diagonal Majorana mass matrix 
which leads to new solutions viable with data. In a basis where 
the Majorana mass matrix is diagonal, 
these solutions imply a Dirac matrix with specific relationships 
amongst its elements.  We find that at the level of total 4 zeros 
together in $m_D$ and $M_R$ 
the mass matrices are almost consistent
with the data but one mixing angle is predicted to be unsuitable.
At the next level, i.e. with total 3 zeros, only seven patterns
of mass matrices describe the experimental data well.
The seven solutions have testable predictions for  the future neutrino
experiments.
In particular, each solution has definite predictions about
the observation of the 1-3 leptonic mixing  angle and the
effective mass measured in neutrino-less double beta decay.
The solutions of the mass matrices contain novel texture forms
and provide new insights into the lepton-generation structure.
We also discuss possible connections between these textures
and the tri-bimaximal mixing to search for symmetry principles
behind the mass matrix structure.
\end{abstract} 

\end{titlepage}

\newpage
\section{Introduction}
Considerable progress in our understanding of neutrino properties 
have been made in the last decade. 
Spectacular
results from neutrino oscillation experiments have established beyond
doubt that neutrinos have mass and they mix \cite{maltonireview}.
For three neutrino generations 
the neutrino mass matrix at low energy is characterized 
by  9 parameters -- the three masses, 
the three mixing angles and the three phases. 
The mixing matrix,  usually known as the Pontecorvo-Maki-Nakagawa-Sakata 
(PMNS) matrix, is expressed in the standard parametrization as 
\be
V =
\left(
\begin{matrix}
c_{12} c_{13} & s_{12}c_{13} & s_{13} e^{-i\delta} \cr
-s_{12}c_{23}-c_{12}s_{23}s_{13}e^{i\delta}
& c_{12}c_{23}-s_{12}s_{23}s_{13}e^{i\delta} & s_{23}c_{13} \cr
s_{12}s_{23}-c_{12}c_{23}s_{13}e^{i\delta}
&-c_{12}s_{23}-s_{12}c_{23}s_{13}e^{i\delta} & c_{23}c_{13}
\end{matrix}
\right),
\label{pmns}
\ee
where $c_{ij}$ and $s_{ij}$ stand for $\cos\theta_{ij}$ and $\sin\theta_{ij}$.
This matrix is to be multiplied from right by a diagonal phase matrix 
$P = {\rm diag}(1, e^{-i \rho/2}, e^{-i \sigma/2})$ where $\rho$ and $\sigma$ 
denote the Majorana phases,
which disappears if the neutrinos are Dirac particles. 
Oscillation experiments, so far have determined
the two mass squared differences and two mixing angles and have 
provided an upper bound on the third mixing angle. 
The current data specify the 3$\sigma$ values of the oscillation parameters 
as presented in Table 1 
\cite{Maltoni:2004ei}. 

Thus the data show that there are two independent mass scales 
with $\Delta m^2_{21}/|\Delta m^2_{31}| = 0.032$ at the best-fit. 
Unlike quark sector where all mixing angles are small in the 
neutrino  sector 
there are  
two large mixing angles while the third one can be small. 
The solar neutrino data have established  that $\Delta m^2 _{21} > 0$. 
But the sign of the atmospheric mass scale $\Delta m^2_{31}$ is not 
yet known. According to the sign of $\Delta m^2_{31}$    
the neutrino spectrum can have two hierarchies, normal 
hierarchy: $m_3^2 \simeq  \Delta m^2_{31} \gg m_2^2 \simeq \Delta m^2_{21} 
\gg m_1^2$ with $\Delta m^2_{31} >0$ or 
inverted hierarchy: $m_2^2 \simeq  m_1^2 \simeq |\Delta m^2_{31}| \gg m_3^2$ 
with $\Delta m^2_{31} <0$. 
The three neutrinos  can also be quasi-degenerate
with $m_3^2 \simeq m_2^2 \simeq m_1^2  \gg \dma$
in which case there is no hierarchy. But one can still ask
what the sign of
$\Delta m^2_{31}$  is. 
Inverted hierarchy and quasi-degeneracy are very unlike to what 
is found in the quark sector. 
Even for normal hierarchy the mass ratio is much weaker than that
in quark sector. 

While, oscillation experiments can determine the mass squared differences 
and the mixing angles,  
information  on absolute neutrino masses can come from 
tritium beta decay or neutrino-less double beta decay. 
The former gives the most direct bound on absolute neutrino masses from 
kinematics and the present bound is $m_\beta < 2.3$ eV (95\% C.L.) coming 
from the Mainz tritium beta decay experiment \cite{mainz}. 
In the standard parametrization (\ref{pmns}), 
$m_\beta$
can be expressed as  
\be \label{eq:KATRIN}
m_\beta = (c_{12}^2 c_{13}^2 m_1^2 + s_{12}^2 c_{13}^2 m_2^2 + s_{13}^2 m_3^2)
^{1/2} ~.
\ee 
Neutrino-less double beta decay violates Lepton number by two units 
and can occur if neutrinos are Majorana particles \cite{scheval}.
The best current limit on the effective mass, which is the 
absolute value of the $ee$ element of the mass matrix, $m_{ee}$ :
\be \label{eq:meff}
m_{ee} = c_{13}^2 c_{12}^2 m_1 + 
e^{i \rho} c_{13}^2 s_{12}^2 m_2 
+ e^{i (\sigma + 2 \delta)} s_{13}^2 m_3
\ee
is given by measurements of
$^{76}$Ge by the Heidelberg-Moscow and IGEX collaboration \cite{HM,IGEX}
\be \label{eq:current}
|m_{ee}| \le 0.35 \, \zeta~{\rm eV}~,
\ee
where $\zeta={\cal O}(1)$ denotes the  uncertainty
coming from the nuclear physics involved in calculating the decay
width of \obb.

\begin{table}
\begin{center}
\begin{tabular}{c|c|c}\hline\hline
& best fit & $3\sigma$ range   \\\hline
$\Delta m_{21}^2$ [$10^{-5}{\rm eV}^2$]  & 7.6 & 7.1 - 8.3 \\
$|\Delta m_{31}^2|$ [$10^{-3}{\rm eV}^2$]  & 2.4 & 2.0 - 2.8 \\\hline
$\sin^2\theta_{12}$ & 0.32 & 0.26 - 0.40 \\
$\sin^2\theta_{23}$ & 0.50 & 0.34 - 0.67 \\
$\sin^2\theta_{13}$ & 0.007 & $\leq$ 0.05 \\\hline\hline
\end{tabular}
\caption{The present best-fit values and the 3$\sigma$ ranges 
of oscillation parameters from \cite{Maltoni:2004ei}.} 
\end{center}
\end{table}

Non-zero neutrino masses and mixing imply physics beyond the 
standard model. 
The most popular mechanism 
for giving small neutrino masses is the seesaw mechanism in which 
one adds heavy right-handed singlets (Type-I)
\cite{seesaw1, Mohapatra:1979ia},
scalar triplets (Type-II) \cite{seesaw2} 
or  fermion triplets (type-III) \cite{Foot:1988aq}
to generate small neutrino masses at low scale. 
In the context of the type-I seesaw mechanism  the light neutrino 
mass matrix is given as
$\mathcal{M} = -m_D M_R^{-1} m_D^{\rm T}$, where $m_D$ is the 
Dirac mass matrix and $M_R$ is the Majorana mass matrix of 
the heavy right-handed neutrinos. 
Apart from the effective neutrino mass matrix $\mathcal{M}$, 
the low energy Lagrangian of the lepton sector also 
contains the charged-lepton mass matrix $M_l$.  
The lepton flavor mixing matrix is defined by 
the two unitary matrices which diagonalize each mass matrix:
$V \equiv  V_l^\dagger V_\nu$. 
If one assumes the charged-lepton mass matrix is diagonal 
then $V_l =1$.  

One way to understand the form  of neutrino masses is through texture zeros
in the Majorana mass matrix  at the low scale 
\cite{Frampton:2002yf,mnuzero}.  
By texture zero we mean those entries
which are vanishingly small as compared to some other elements.
This approach had been adopted in quark sector \cite{quarkM}
and therefore it seems
plausible that this may work for the lepton sector also. 
The origin of such zero entries could be traced to symmetry or
dynamics lying behind the Yukawa sector of the standard model,
for example, the $U(1)$ symmetry by Froggatt and Neilsen
\cite{Froggatt:1978nt} with supersymmetry, 
or other flavor symmetries \cite{models} which include 
either discrete or continuous groups. 
The stability of texture zeros in $\mathcal{M}$ against renormalization group 
effects have been studied in \cite{mnuRGE}. 

Within the framework of the seesaw mechanism it is often 
considered more natural to study texture zeros appearing in 
the Yukawa coupling matrix $m_D$  and/or the right-handed 
Majorana mass matrix $M_R$ 
\cite{seesawTEX}.
In general the seesaw framework contains more parameters compared to 
what can be obtained from measurements at low energy and 
it is not possible to fix the high energy parameters entirely from 
low energy data. 
Texture zeros in the matrices $m_D$ and/or $M_R$ 
can help in reducing the number of parameters, and thus 
strengthen the predictive power of the model. 

Another way to increase the predictability and reduce the number of 
high scale parameters 
of the  seesaw model is to  reduce the number of right-handed neutrinos. 
The minimal number of right-handed neutrinos  with which low energy 
phenomenology compatible with current data  
can be obtained is two \cite{2nuR}.
With one heavy neutrino after seesaw diagonalization 
the mass matrix at low scale is rank 1, that is,  it contains 
two zero eigenvalues and hence is not consistent with the current data. 
  
In this paper we couple the two ideas and do an extensive and 
systematic analysis of all possible texture zeros in $m_D$ 
and $M_R$ in the framework of the minimal seesaw 
model containing two heavy right-handed neutrinos. 
There already exists exhaustive analysis of possible 
texture zeros in $m_D$ in the literature in the context of 
the minimal seesaw model \cite{2nurTEX}.
But most analyses considered a diagonal form for the 
Majorana mass matrix $M_R$.  
The possibility of a non-diagonal $M_R$ and related constraints 
on seesaw parameters have been discussed in few papers but this 
discussion is not exhaustive. 
In addition, it is not apparent that any mechanism leading to zeros 
in $m_D$ will necessarily require a diagonal form of $M_R$.
It is thus worthwhile to take into account the full generality
of $M_R$ and perform an exhaustive classification of the textures 
according to the total number of zeros of $m_D$ and $M_R$ together. 
By this systematic analysis, we encounter not only the textures 
which have been discussed in the literatures \cite{seesawTEX, 2nuR,2nurTEX}
but also new viable textures arising out of the general treatment
for $M_R$,  which have not been discussed.

It is important to keep in mind that, in a texture analysis,
one should not fix the basis of the generation space in advance
While it is true that 
one can always move from one 
basis to another by unitary transformations acting on the fields,
without  changing any physical consequences,
(for example, from some general basis to the one in which $M_R$ is 
diagonal by redefining the right-handed neutrinos)
this does not mean that one can always find full
possibilities of textures by examination of the texture zeros
in the $M_R$ diagonal basis.
The unitary transformation for fields is just a change of the 
coordinate which describes a physical system, whereas the different
(or independent) textures correspond to  different systems,
namely, different Lagrangians.
A texture zero in some basis can appear as definite relations 
among matrix elements in other basis.
For example, some texture zeros of $m_D$ and $M_R$ in $M_R$ non-diagonal 
basis is mapped to the specific  relations among mass matrix elements
of $m_D$ 
in $M_R$ diagonal basis.
Thus the texture zeros are "hidden" in $M_R$ diagonal basis,
in the sense that one cannot reach the texture forms of
the general basis as long as one merely impose 
vanishing elements in $m_D$.

We classify the textures by clarifying whether
each combination of textures can fit the experimental results or not.
This is the only principle we will take.
In particular, we do not stand on any aesthetic discussion
and as long as a certain texture can fit the data,
we regard it as viable even if it needs fine tuning of parameters
to reproduce the observable.
Since a priori, we do not know what kind of symmetry is lying
behind some fine tuning,
we believe this thorough analysis might bring useful results
towards deeper understanding for the generation structure.

Throughout the analysis we assume that the charged-lepton mass matrix
is diagonal. 
Because of the observed mass hierarchy of the charged-lepton masses,
it is likely that the contribution to the PMNS matrix is small from 
the point of view of grand unification, though there are
interesting possibilities of highly asymmetric forms of the charged-lepton 
mass matrix \cite{lopsided}. 
In this paper we do not consider these possibilities and simply assume that 
the neutrino sector is responsible solely for the structure of the lepton 
generation mixing. 

The plan of the paper goes as follows. In the next section we discuss the 
rudimentary features of the seesaw model with two right-handed 
neutrinos. 
The following section discusses the number of possible texture 
zeros in the Dirac matrix $m_D$ and the Majorana mass matrix $M_R$
and then does a detailed classification of the patterns based on 
the number of texture zeros in $m_D$ and $M_R$ together. 
For each pattern we discuss the compatibility with the current 
oscillation data, prediction for the 1-3 leptonic mixing angle and 
the effective mass parameter constrained by neutrino-less double 
beta decay.   
We end in Section 4 by summarizing our results. 

\section{The model with two right-handed neutrinos} 

The leptonic part of the 
Yukawa interactions in presence of three left-handed and two 
right-handed neutrinos can be written as 
\be
-{\cal{L}} \,=\, (Y_{\nu})_{ij} \overline {N_{R_j}}  \, \tilde{\phi}^\dag 
l_{L_i} \,+\, 
(Y_l)_{ij} \overline{E_{R_j}} \phi^\dagger l_{L_i}  
\,+\, 
\frac{1}{2} \overline{{N_{R_i}}^c} (M_R)_{ij} N_{R_j} \,+\, {\rm h.c},  
\ee
where $\phi$ denotes $SU(2)$ higgs doublet with 
$\tilde{\phi} = i \sigma^2 \phi^{*}$, 
the lepton doublet of flavor $i$ is denoted by $l_{L_i}$, 
$E_{R_i}$ are the right-handed charged-lepton singlets and 
$N_{R_i}$ denote the right-handed neutrino fields which are singlets
under the standard model gauge group. 
The Yukawa coupling constants 
$Y_\nu$ and $Y_l$ are complex-valued $3\times 3$ matrices.
After the electroweak symmetry breaking one gets the 
charged-lepton mass matrix $M_l = v Y_l$ and the Dirac mass matrix 
for the neutrino as $m_D = v Y_\nu$ where $v$ is the vacuum expectation 
value of the neutral component of the higgs doublet $\phi$.    
The Majorana mass matrix $M_R$ is $2 \times 2$ complex symmetric
matrix.
The mass matrix for the neutral fermions can be written as 
\begin{eqnarray}
M_\nu = \begin{pmatrix}
0 & m_D \\
m_D^{\rm T} & M_R \\
\end{pmatrix}.
\label{massmatrix}
\end{eqnarray}
The light neutrino mass matrix after the seesaw 
diagonalization assuming $M_R \gg m_D$ is given by 
\be
\mathcal{M} = -m_D M_R^{-1} m_D^{\rm T}.  
\label{seesawF}
\ee
Since $M_R$ can be high the mass eigenvalues are naturally 
suppressed. 
In the 3+2 model $m_D$ is a $3 \times2$ matrix and $M_R$ is a $2 \times 2$ 
matrix while the light neutrino mass matrix $\mathcal{M}$ is $3 \times 3$. 
We note that the 3+2 model can be considered as the limiting case of a 
3+3 model where the heaviest neutrino is extremely heavy
compared to the others so that the contribution of the heaviest
neutrino is negligibly small.
In general for the $3 \times 2$ case there will be 
9 free parameters characterizing the Yukawa matrix $m_D$ and  4 parameters
for $M_R$ giving a total of 13 free parameters.  
Thus there is already a reduction from 24 to 13 as compared to the 3+3 
model. As we will see later for the cases of texture zeros in $m_D$ and 
$M_R$ the number of free parameters can be reduced even further.  

In general the Majorana mass matrix $M_R$ is non-diagonal
in the basis where the charged current is flavor diagonal. 
It is thus written as
\be
 U_R^\dagger M_R U_R^* = {\rm diag}(M_1, M_2).
\label{mrdiagbasis}
\ee
One can make a basis rotation 
so that  the right-handed Majorana 
mass matrix $M_R$ becomes diagonal by the unitary matrix $U_R$. 
However in that case the Dirac mass 
matrix $m_D$ also gets modified to  $m_D U_R^{*}$. 
Let us denote the modified Dirac mass matrix as
\begin{eqnarray}
m_D U_R^{*} \,\equiv\, 
\left( {\bf a}_1 , {\bf a}_2 \right),
\end{eqnarray}
where ${\bf a}_1$ and ${\bf a}_2$ are column vectors which have three 
elements.
With this notation we can write the seesaw formula (\ref{seesawF})
as
\begin{eqnarray}
\mathcal{M} \,=\,
- \sum_{i = 1}^{2} {\bf a}_i\cdot {\bf a}_i^{\rm T}\frac{1}{M_i}.
\end{eqnarray}
Note that the summation is stopped at 2 because we assume only
two right-handed neutrinos take part in the seesaw mechanism.
This expression highlights the most prominent feature of the seesaw 
mechanism with two heavy neutrinos. 
That is, the rank of the induced Majorana mass matrix is at most 2,
which means that we have at least one massless left-handed neutrino.
Since the renormalization group running does not affect the
rank of the mass matrix, it is a scale independent feature
of the two right-handed seesaw framework.

The Majorana mass matrix $\mathcal{M}$ is symmetric and
can in general be diagonalized as  
\be
V_\nu^{\rm T} \mathcal{M} V_\nu = D_\nu
\ee
where $D_\nu$ is the diagonal (real and positive) mass eigenvalues 
for the left-handed neutrinos: $D_\nu= {\rm diag}(m_1, m_2, m_3)$, 
and $V_\nu$ is an unitary matrix which includes 3 angles and 6 phases 
in general. 
The charged-lepton mass matrix $M_l$ is diagonalized as
\begin{eqnarray}
V_l^{\rm T} M_l U_l^* \, = \, D_l,
\end{eqnarray}
where $V_l$ and $U_l$ are unitary matrices which can be removed
by unitary rotations of the left and the right-handed charged-leptons.
The diagonal matrix $D_l$ denotes the observed charged-lepton masses:
$D_l = {\rm diag}(m_e, m_\mu, m_\tau)$.
The generation mixing for leptons is described by the
Pontecorvo-Maki-Nakagawa-Sakata (PMNS) matrix, which is defined by
the product of the two unitary matrices;
\begin{eqnarray}
V \equiv V_l^\dag V_\nu.
\end{eqnarray}
A general $3\times 3$ unitary matrix can be parameterized
by three 3 angles and 6 phases.
Out of 6 phases, 3 phases can be removed by the redefinition
of the left-handed neutrino fields.
Thus 3 angles and 3 phases can affect observables if the
neutrinos are Majorana particle.
In the case where the neutrinos are Dirac particle,
we can further remove 2 phases by using the right-handed neutrinos,
and there is only 1 phase which is responsible for CP violation,
just as in the quark sector.
  
In the following discussion, we assume
that the charged-lepton mass matrix is diagonal, so that
$V_l = I$. 
While this treatment will not cover whole possible solutions
in the lepton sector, it turns out that this is indeed a good 
first step to extract physics involved in the seesaw mass matrices 
with two right-handed neutrinos.
Against the diagonal charged-lepton mass matrix, we will perform
texture analysis step by step from the most minimal (maximal number
of zero) level.
We discuss not only successful textures, but also the textures
which are not totally compatible with the experimental data.
These textures can become viable with the inclusion of non-diagonal 
charged lepton matrices and from an understanding of which sector 
is inconsistent with data it will be possible to determine the 
form of the 
non-diagonal charged-lepton textures 
which are needed to make the   
``close to viable'' seesaw texture acceptable. 

Finally, we would like to note that the generation
indices for the lepton doublet is thus fixed in any discussions below.
In particular, it should be noted that any exchanging operation for 
the rows of $m_D$ do affect physical consequences (however small they are)
so that the textures which are related to each other by such exchange
should be regarded as independent textures.

\section{Texture analysis} 
In this section we will first consider the texture zeros of 
$m_D$ and $M_R$ separately and then check the total  number of 
texture 
zeros together in $m_D$ and $M_R$. 
Here the number of the zero means the number of the independent
vanishing elements in each matrix.
For example, for a symmetric matrix, ``1 zero'' means
that a diagonal element or a pair of off-diagonal components
in symmetric positions are anomalously small compared to the other
elements.

\subsection{Zeros of the Dirac mass matrix $m_D$}
\label{Dirac}
First of all, let us consider the minimality of $m_D$, without
taking into account the form of $M_R$. 
The following cases may arise: 
\begin{itemize}

\item{\bf{More than 3 zeros}} 

It should be useful to note that if the number of vanishing elements
in $m_D$ is $\geq$ 4, it cannot lead to viable forms of the effective 
Majorana matrix $\mathcal{M}$.
This is because if we have 4 or more than 4 zeros in $m_D$,
there is at least one vanishing row in the Dirac mass matrix.
This means that at least one left-handed neutrino is decoupled
from the right-handed states so that the neutrinos can be mixed
only between the other two states which are coupled with the
right-handed neutrinos.
We can thus exclude more than 3 zeros in $m_D$.

\item{\bf{3 zeros}} 
 
The next possibility is to consider  
three zeros in $m_D$. 
We have ${}^6{\rm C}_3 = 20$ patterns of matrices as the general
possibilities of 3 zero textures in $m_D$.
They can be classified into three categories.
The first one includes the matrices which have one vanishing row.
There are 12 patterns of such matrices but none of them can
reproduce observation, as we saw in the discussion above.
The second one includes the matrices which have one vanishing column.
The 2 patterns of such matrices are
 also not viable because
the vanishing column implies that only one right-handed neutrino
takes part in the seesaw mechanism, which leads to two
massless states. 
The third category is formed by the other 6 patterns.
We cannot exclude these patterns as long as we are concerned with 
only Dirac mass matrix $m_D$.
An example in this category is
\begin{eqnarray}
m_D = \begin{pmatrix}
0 & d \\
b & 0 \\
c & 0 \\
\end{pmatrix}.
\label{3zeromD}
\end{eqnarray}
The other 5 patterns are obtained by permuting the rows and columns
of (\ref{3zeromD}).
In the following discussions, we will examine these 6 patterns of $m_D$
as general possibilities of the 3 zero $m_D$.

\item{\bf{2 zeros}} 

The next to minimal number of zeros is 2.
In this case we have ${}^6{\rm C}_2 = 15$ patterns of matrices as
the general possibilities.
They can be classified into three categories.
The first one includes the 3 matrices which have one vanishing row.
As we saw in the discussion above, we need not to examine these three.
The second one includes the 6 matrices which have two zero entries
in the same column, for example
\begin{eqnarray}
m_D = \begin{pmatrix}
a & 0 \\
b & 0 \\
c & f \\
\end{pmatrix}.
\label{2zeromDa}
\end{eqnarray}
The other five patterns are obtained by permuting the rows and the
columns of (\ref{2zeromDa}).
The third category is formed by the other 6 patterns, which
have two zero elements in different columns.
For example,
\begin{eqnarray}
m_D = \begin{pmatrix}
a & 0 \\
0 & e \\
c & f \\
\end{pmatrix}.
\label{2zeromDb}
\end{eqnarray}
The other five patterns are obtained by permuting the rows and the
columns of (\ref{2zeromDb}) (or the permutations of the rows only).
The 12 matrices which belong to the last two categories
are not excluded a priori.
In the following discussions, we take these 12 patterns
as general possibilities of the 2 zero $m_D$.
\end{itemize}

\subsection{Zeros of $M_R$} 
\label{Majorana}
Since we have a $2 \times 2 $ $M_R$ and we include the possibility of 
non-diagonal $M_R$ we can have the following options 
\begin{itemize} 
\item {\bf{ 3 zeros}} 

This gives a $M_R$ with all entries as zero and so this is excluded. 

\item{\bf{ 2 zeros}}

In this case there are two possibilities \\

(i) The diagonal entries are zero 
\begin{eqnarray}
M_R = \begin{pmatrix}
0 & M_{12} \\
M_{12} & 0 \\
\end{pmatrix}.
\label{2zeroMRa}
\end{eqnarray}

(ii) One diagonal entry and the off-diagonal entries are zero 
\\
\begin{eqnarray}
M_R = \begin{pmatrix}
0 & 0  \\
0 & M_{22} \\
\end{pmatrix}.
\label{2zeroMRb}
\end{eqnarray}

or 

\begin{eqnarray}
M_R = \begin{pmatrix}
M_{11} & 0  \\
0 & 0 \\
\end{pmatrix}.
\label{2zeroMRc}
\end{eqnarray}
The last two options give vanishing determinants and give rise to 
a state which does not receive seesaw suppression in mass.
In this paper, we do not consider such exotic spectrum
though it is an interesting possibility to accommodate more than 
two mass differences.
We therefore conclude that only the option (i) can be viable. 

\item{\bf{1 zero}}

The possible options for this case are 
\begin{eqnarray}
M_R = \begin{pmatrix}
0 & M_{12} \\
M_{12} & M_{22} \\
\end{pmatrix}.
\label{1zeroMRa}
\end{eqnarray}
\begin{eqnarray}
M_R = \begin{pmatrix}
M_{11} & M_{12} \\
M_{12} & 0  \\
\end{pmatrix}.
\label{1zeroMRa}
\end{eqnarray}
\begin{eqnarray}
M_R = \begin{pmatrix}
M_{11} & 0 \\
0 & M_{22}  \\
\end{pmatrix}.
\label{1zeroMRa}
\end{eqnarray}
All of these matrices have non-vanishing determinant. 
We thus regard these three matrices as general possibilities 
in the following analysis.

\end{itemize} 

\subsection{Zeros of $m_D$ and $M_R$ combined -- almost viable forms --} 
So far we have discussed the texture zeros for the Dirac mass matrix
$m_D$ and the Majorana mass matrix $M_R$ separately.
In this section we search for the combinations of
$m_D$ and $M_R$ which are compatible with the present oscillation data,
keeping the results of Section \ref{Dirac} and \ref{Majorana} in mind.
We classify the combinations of the two matrices according to total  
number of texture zeros to be distributed in  $m_D$ and $M_R$. 
We will find that at the 4 zero level, several textures appear
to be close to perfectly viable.

Let us consider total 6 zero elements in the mass matrices
$m_D$ and $M_R$ as a starting point of texture analysis.
The maximum number of texture zeros admissible in $m_D$ is three. 
This leaves us with the rest of the three zeros in $M_R$ which is 
excluded. 
Thus there are no viable textures in the case of the total 6 zeros.

We next consider the possibility of having total 5 zeros in $m_D$ and $M_R$.
According to the discussion in Section \ref{Dirac} and \ref{Majorana},
we can only have the case where $m_D$ has 3 zeros and $M_R$ has 2 zeros.
Then $\mathcal{M}$ is given as,  
\begin{eqnarray}
\mathcal{M} 
&=& \begin{pmatrix}
0 & d \\
b & 0 \\
c & 0 \\
\end{pmatrix}
\begin{pmatrix}
0 & \frac{1}{M_{12}} \\
\frac{1}{M_{12}} & 0 \\
\end{pmatrix}
\begin{pmatrix}
0 & b & c \\
d & 0 & 0 \\
\end{pmatrix}\nonumber\\
&=& \begin{pmatrix}
0 & bd & cd \\
bd & 0 & 0 \\
cd & 0 & 0 \\
\end{pmatrix}\frac{1}{M_{12}}.
\label{5zeroEG}
\end{eqnarray}
It should be noted that $M_R$ is invariant under the exchange
of the generation label.
By virtue of this feature, the resultant Majorana mass matrices of
the other possible combinations are obtained by
permuting the rows and the columns of (\ref{5zeroEG}).
It is interesting to note that the texture form (3.10) can be obtained
from $ L_e - L_\mu - L_\tau$ flavor symmetry.
While naive bi-maximal structure can be produced with inverted hierarchy, 
it is clear that the two nonzero masses are degenerate.
Thus we conclude that there is no solution in the level of the total 5 zero
texture.

The next step is the total 4 zero textures.
In this case, there are two possibilities;  
\begin{enumerate}
\item[(i)]  3 zeros in $m_D$ and 1 zero in $M_R$
\item[(ii)]  2 zeros in $m_D$ and 2 zeros in $M_R$
\end{enumerate}
In the following we investigate the above cases in detail one by one.

\bigskip
\begin{flushleft}
\begin{framebox}
{(i)~~{\bf3 zero $m_D$ and 1 zero $M_R$}}
\end{framebox}
\end{flushleft}
\bigskip
To study this case, it is convenient to write down
an example of the seesaw formula in terms of a most general $2 \times 2$ 
$M_R$ as 
\begin{eqnarray}
\mathcal{M} &=&
\begin{pmatrix}
0 & d \\
b & 0 \\
c & 0 \\
\end{pmatrix}
\begin{pmatrix}
A &B  \\
B &C  \\
\end{pmatrix}
\begin{pmatrix}
0 & b & c \\
d & 0 & 0 \\
\end{pmatrix}\nonumber\\
&=& \begin{pmatrix}
d^2 C & bd B & cd B\\
bdB & b^2 A & bc A \\
cdB & bcA & c^2A \\
\end{pmatrix},
\label{aa}
\end{eqnarray}
where we introduce the parameters $A,B,C$ just to simplify the
notation:
\begin{eqnarray}
\begin{pmatrix}
A & B \\
B & C \\
\end{pmatrix}
\equiv (M_R)^{-1} = \frac{1}{M_{11}M_{22} - M_{12}^2}
\begin{pmatrix}
M_{22} & -M_{12} \\
-M_{12} &M_{11}\\
\end{pmatrix}.
\end{eqnarray}
In the following, we examine 1 zero $M_R$ by taking
$A$, $B$ or $C$ to be zero in turn.
As we saw in Section \ref{Dirac},
there are 6 patterns of 3 zero $m_D$ which we
should consider.
Although (\ref{aa}) presents just an example of the 
6 possible combinations of 3 zero $m_D$ and general $M_R$,
we can study the consequences of the other 5 patterns
by careful observation of (\ref{aa}).

Keeping $A$, $B$, and $C$ to be non-zero, 
we have 12 real parameters characterizing $m_D$ 
and $M_R$. All the 3 phases of $m_D$ and 2 phases of $M_R$ 
can be removed by redefining the fields and so for this case 
we have 7 parameters -- 3 real parameters for $m_D$ and 4 real 
parameters for $M_R$ (one of which is a phase).
With no loss of generality, we can take the basis in which
the matrix element $C$ has a phase. 
We thus regard only $\mathcal{M}_{11}$ as complex in what follows. 

The most striking feature of the matrix (\ref{aa}) is that
it has an eigenvector which is associated with the zero eigenvalue;
that is, $(0,\,-c/\sqrt{b^2 + c^2}\, ,\, b/\sqrt{b^2 + c^2}\,)^{\rm T}$.
Note that this vector does not depend on the phase of $\mathcal{M}_{11}$.
Since the zero element in this eigenvector must be interpreted
as the reactor angle, the mass ordering of the neutrinos is fixed
to be the inverted hierarchy.
Since this matrix yields a vanishing $\theta_{13}$ it is characterized 
by a class of the discrete $Z_2$ symmetry 
\cite{Grimus:2004cc}.
The atmospheric angle is controlled by the elements $b$ and $c$. 
By taking $ b \simeq\!\!(=) c$, we have $\theta_{23} \simeq\!\!(=)45^{\circ}$.
This corresponds to $\mu-\tau$ exchange symmetry. 

We have three options to take 1 zero texture in $M_R$,
that is, $A= 0$ , $B=0$ or $C=0$.
In these options, we find that $C = 0$ is
the most promising texture and the other two cases are
disfavored at the first glance:
\begin{itemize}

\item $B=0$:
This case corresponds to a 3 zero $m_D$ with a diagonal $M_R$. 
In this case, there is only one mixing -- in the 2-3 sector 
and thus cannot reproduce two large mixing angles. 

\item $A=0$: There are three zeros  in the Majorana mass matrix 
$\mathcal{M}$ which is not viable with current data \cite{Frampton:2002yf}.  
We note that in this case the
1-2 mixing angle will be given as 
\begin{eqnarray}
\sin\theta_{12} 
& \simeq & \frac{1}{\sqrt{2}} \,+\, \frac{1}{8\sqrt{2}}\alpha \,\,,
\end{eqnarray}
where $\alpha$ is the small parameter
$ \alpha \,\equiv\, \Delta m_{21}^2/|\Delta m_{31}^2| $.
Thus the 1-2 angle is greater than the maximal value of $\pi/4$ which is 
ruled out by solar neutrino data \cite{solarmaximal}. 

\end{itemize}
Note that the logic which we have used to exclude $A=0$ and $B=0$
are also applicable to the case of the other 5 patterns
of $m_D$ because the Majorana mass matrix $\mathcal{M}$ for
the other 5 textures can be obtained by permuting rows and columns
of (\ref{aa}). 

By setting $C=0$, we lose the sole matrix element which is responsible for
the CP violation.
This means that there is no CP violating phase in the lepton Yukawa
Lagrangian.
As we have seen the above, there is an eigenvector
$(0,\,-c/\sqrt{b^2 + c^2}\, ,\, b/\sqrt{b^2 + c^2}\,)^{\rm T}$.
This fact forces the mass spectrum to be the inverted hierarchy.
Moreover, we can see that it needs $b \simeq c$ to produce the 
observation of the atmospheric neutrinos.
The inverted mass spectrum is then realized in the region
where $\mathcal{M}_{12} \simeq \mathcal{M}_{13}  \gg 
\mathcal{M}_{22} \simeq \mathcal{M}_{23} \simeq \mathcal{M}_{33}$ is satisfied.
The magnitudes of the two nonzero mass eigenvalues $m_1$ and $m_2$
are controlled by the magnitude of 
$\mathcal{M}_{12} \simeq \mathcal{M}_{13}$
and the mass difference between these two states are ruled by 
$\mathcal{M}_{22} \simeq \mathcal{M}_{23} \simeq \mathcal{M}_{33}$.
This means that the 1-2 rotation angle is controlled by the ratio 
$dB/bA$ in such a way that $\tan 2\theta_{12} \simeq 2dB/bA \gg 1$.
Thus the solar angle will be predicted as $\simeq 45^{\circ}$.
In fact, we can write down the solar angle $\sin\theta_{12}$
in terms of the mass differences as
\begin{eqnarray}
\sin\theta_{12}
&\simeq& \frac{1}{\sqrt{2}}
-\frac{1}{8\sqrt{2}}\alpha.
\label{v12pre}
\end{eqnarray}
The prediction (\ref{v12pre}) is due to the absence of the
1-1 matrix element in the low energy Majorana mass matrix.
Although the solar angle is predicted to be smaller than $45^\circ$,
the deviation from maximal is too small to be compatible with the data.

As for the other possible combinations, we can see the consequences
immediately from the above one example.
As we saw in Section \ref{Dirac},
there are 6 patterns of 3 zero $m_D$ which we
should consider, and they are related each other by permutations of the rows
and the columns.
The one of them is just the texture which we took in (\ref{aa}).
Thus the other 5 resultant Majorana mass matrices are obtained by
exchanging the rows and the columns of (\ref{aa}), and replacing
the matrix element $A$ with $C$.
In the case of the 2 patterns of $m_D$ which are obtained by permuting
rows of $m_D$ in (\ref{aa}), it should be noted that we cannot replace
the first row with the other one, otherwise the zero element in the
mixing matrix comes in the  wrong position.
This means that these  2 patterns can be safely excluded.

As for the 3 patterns which are associated with exchanging the
columns of $m_D$ in (\ref{aa}), we can discuss in the same manner
as the above by replacing $A$ with $C$.
We find that the Dirac mass matrix obtained by a column exchange 
\begin{eqnarray}
m_D = \begin{pmatrix}
d & 0 \\ 0 & b \\ 0 & c \\
\end{pmatrix}
\label{aa2}
\end{eqnarray}
is the most promising one.
The Majorana mass matrix $\mathcal{M}$ for this case can be related to 
the general case considered in (\ref{aa})
by exchange of label of the right-handed neutrinos 
$ A \leftrightarrow C$ {\it i.e.} 
\begin{eqnarray}
M_R^{-1} = \begin{pmatrix}
C & B \\ B & A \\
\end{pmatrix}
\label{mrinv2}
\end{eqnarray}
It is clear from (\ref{aa}) that after the exchange $ A \leftrightarrow C $
the promising texture in this case is obtained by putting A=0. 
In summary, in the case of 3 zero $m_D$ and 1 zero $M_R$,
we found that the most promising texture, which can give one 
zero and two large mixing angles, is the $C=0$ case in
(\ref{aa}) and  $A=0$ with (\ref{aa2}).
However, these two combinations are related each other by the
label exchange of the two heavy neutrinos so that these two
solutions describe exactly the same physics.
Over all, we conclude $C=0$ case in (\ref{aa}) 
(or $A=0$ with (\ref{aa2})) is the most promising texture.
The predictions are very distinctive;
\begin{itemize}

\item The inverted hierarchy with $m_3 = 0$.
\item $\theta_{13} = 0$.
\item No $\cancel{\rm CP}$ at all (at high energy as well as low energy).
\item $\sin\theta_{12} \simeq \frac{1}{\sqrt{2}}
-\frac{1}{8\sqrt{2}}\alpha.$

\end{itemize}
The atmospheric angle $\theta_{23}$ is well controlled, and
we can reproduce the best fit value $\theta_{23} = 45^{\circ}$.
Thus this texture, which can be obtained since we have gone beyond 
the assumption of a diagonal $M_R$, 
needs some correction to the 1-2 mixing angle.
Such a correction can be traced to various sources.
For example, it is clear that 1-2 mixing from charged-lepton sector can
ameliorate the problem immediately.
As another example, it might be possible to cure the problem
by the renormalization group effect from some (high)
energy scale (at which the texture zeros are imposed) to the electroweak
scale \cite{lmarg}.

\bigskip
\begin{flushleft}
\begin{framebox}
{(i)~~{\bf2 zero $m_D$ and 2 zero $M_R$}}
\end{framebox}
\end{flushleft}
\bigskip
Next let us explore the case of 2 zero $m_D$ and 2 zero $M_R$.
As we have discussed in Section \ref{Majorana},
for two zero $M_R$ the only allowed possibility with non-vanishing 
determinant is the one with zeros at the diagonal positions.  
Note that in a basis where $M_R$ is diagonal, the elements of $m_D$ 
in this basis are made out of the linear combinations of the elements
in the original basis.
Therefore texture zeros for $m_D$ in the new basis imply
particular relations between matrix elements of $m_D$ in the old basis,
so that the texture analysis considering zeros of $m_D$ and a diagonal 
$M_R$ would not include this 2 zero $m_D$ and 2 zero $M_R$ case.   

As we saw in Section \ref{Dirac},
we should consider two types of 2 zero $m_D$.
Namely, the 6 patterns represented by (\ref{2zeromDa}) with the
two  zero entries in the same column  and the other 
6 patterns represented by (\ref{2zeromDb}) with the zero entries 
in different columns. 
A representative for the 6 combinations which involve the former
type of $m_D$ is 
\begin{eqnarray}
m_D = \begin{pmatrix}
a & d \\ b & 0 \\ c & 0 \\
\end{pmatrix},\quad
M_R^{-1} = \begin{pmatrix}
0 & B \\ B & 0 \\
\end{pmatrix},
\label{4zeroA}
\end{eqnarray}
However these mass matrices induce the same form of the left-handed
Majorana mass matrix as that of (\ref{aa}) with $A=0$.
We thus drop this combination from the list of promising texture. 
Furthermore, we can also drop the other 5 combinations because
their effective Majorana mass matrices can be obtained by
permuting rows and columns of that of (\ref{4zeroA}).
We thus conclude that there is no viable texture in this category.

For the group represented by (\ref{2zeromDb}), we have an example
\begin{eqnarray}
m_D = \begin{pmatrix}
a & 0 \\ 0 & e \\ c & f \\
\end{pmatrix},\quad
M_R^{-1} = \begin{pmatrix}
0 & B \\ B & 0 \\
\end{pmatrix}.
\label{4zeroB}
\end{eqnarray}
After the seesaw integration, we have
\begin{eqnarray}
\mathcal{M} \,=\,
\begin{pmatrix}
0 & ae & af \\
ae & 0 & ce \\
af & ce& 2cf \\
\end{pmatrix}\!B.
\end{eqnarray}
There remains two texture zeros at the low energy scale.
We note that by the redefinition of the fields, all the phase
degrees of freedom can be moved away from the Yukawa Lagrangian
of the lepton sector.
Thus there is no CP violating phenomena with the texture (\ref{4zeroB})
at high energy as well as low energy.

The texture combination (\ref{4zeroB}) can accommodate
both the normal and the inverted hierarchy.
We study them individually.
\paragraph{Normal hierarchy}

In the case of the normal hierarchy, the neutrino masses $m_1$, $m_2$
and $m_3$ are fixed as $m_1 = 0$, $m_2 = \sqrt{\Delta m_{21}^2}$
and $m_3 = \sqrt{\Delta m_{31}^2}$.
Since we have only 3 effective parameters after the seesaw
operation, we have two weak eigen-basis invariant predictions;
\begin{eqnarray}
\frac{\tan\theta_{13}}{\sin\theta_{12}}
&=&
\alpha^{\frac{1}{4}},
\vspace{2mm}\label{a1}\\
\sin\theta_{23}
&\simeq&
\alpha^{\frac{1}{4}}
-\frac{1}{2}\alpha^{-\frac{1}{4}}\sin^2\theta_{13},
\label{a2}
\end{eqnarray}
where $\alpha \equiv \Delta m_{21}^2/\Delta m_{31}^2$.
Substituting $3\sigma$ lower bound of 
$\alpha$ and $\sin\theta_{12}$ into (\ref{a1}), we find
\begin{eqnarray}
\sin\theta_{13} \,=\, 0.20.
\label{bound13} 
\end{eqnarray}
Thus with the marginal values of $\alpha$ and $\sin\theta_{12}$ 
we already reach just below the current upper bound 
on $\sin\theta_{13}$ in  (\ref{bound13}).  
Moreover, from the relation (\ref{a2}), we can see that there 
is an anti-correlation between $\sin\theta_{13}$ and $\sin\theta_{23}$ 
in the sense that for a smaller $\sin\theta_{13}$,  $\sin\theta_{23}$ 
will be larger. 
Hence a conservative upper bound of $\sin\theta_{23}$
can be obtained in the limit $\theta_{13} \to 0 $.
Then we find
\begin{eqnarray}
\sin\theta_{23} \,\lesssim \, 
\alpha^{\frac{1}{4}} = 0.42,
\label{a34}
\end{eqnarray}
where we used the best fit values of the mass differences
on the right-hand side.
While the solar and the reactor angles are within present $3\sigma$ data,
we need some corrections to fit the atmospheric angle.

As we saw in \ref{Dirac}, the other 5 patterns of Dirac
mass matrices are obtained by permuting rows of (\ref{2zeromDb}).
It is useful to note that the form of $M_R$ does not change if
we exchange its rows and columns.
Thus, the Dirac textures which are related with column exchange
induce the same predictions and they cannot be independent solution
each other. 
This fact reduces the number of the
mass matrices which we need to examine 5 to 2, that is,
it is enough to consider 2 patterns which is obtained by
exchanging 1-3 and 2-3 rows of (\ref{2zeromDb}).

For the 1-3 exchanging texture, the PMNS matrix element $|V_{33}|$ for
the case of (\ref{2zeromDb})
is identified as the reactor angle.
However, we can see from (\ref{a2}) that this element 
is approximated as
\begin{eqnarray}
|V_{33}| \,\simeq\, 1 - \frac{1}{2}\alpha^{\frac{1}{2}}.
\end{eqnarray}
Thus the proper magnitude for the reactor angle cannot realized
at all.
We conclude that 1-3 exchanging texture from (\ref{2zeromDb}) 
is excluded because of this large discrepancy in $\theta_{13}$.

On the other hand, the 2-3 exchanging texture should be regarded
as the same level as the texture of (\ref{2zeromDb}).
The conservative upper bound (\ref{a34}) implies that
the lower bound of $|V_{33}| \geq 0.82$ from the
normalization condition of the third column vector in the PMNS matrix.
Since the allowed range of the atmospheric angle is almost symmetric 
around the maximal value $45^\circ$,  the deviations from the
best fit value are the same in both textures.

\paragraph{Inverted hierarchy}
In the case of the inverted hierarchy, the masses $m_1$, $m_2$
and $m_3$ are fixed as $m_1 = \sqrt{|\Delta m_{31}^2|}$,
$m_2 = \sqrt{|\Delta m_{31}^2| + \Delta m_{21}^2}$ and $m_3 = 0$.
There are two relations among observables;
\begin{eqnarray}
\sin\theta_{13} 
&\simeq& 
\frac{1}{4 \tan\theta_{23}}\,\alpha,
\label{a38}\\
\nonumber\\
\sin\theta_{12} 
&\simeq& 
\frac{1}{\sqrt{2}}
\,-\, \frac{1}{8\sqrt{2}}\,\alpha 
\,-\, \frac{1}{2\sqrt{2}}\sin\theta_{13},
\label{a39}
\end{eqnarray}
where $\alpha \equiv \Delta m_{21}^2 /|\Delta m_{31}^2| $.
From (\ref{a38}), we can see that $\sin\theta_{13}$ is predicted
to be small;
$\sin\theta_{13} \sim \mathcal{O}(10^{-3})$
(with the best fit values of $\theta_{23}$ and the
mass differences, we have $\sin\theta_{13} = 8.0 \times 10^{-3}$).
As for the solar angle, we can see
from (\ref{a39}) that it becomes near the maximal value
$\theta_{12} \simeq 45^\circ$.
Thus we see that the texture (\ref{4zeroB}) predicts nearly bi-maximal 
mixing with the inverted mass ordering.
The deviation from the bi-maximal form is observed to be small
and it is characterized by the magnitude of the reactor angle
$\sin\theta_{13} \sim \mathcal{O}(10^{-3})$.
Thus this case also needs correction to the 1-2 mixing angle to 
become a perfectly viable.

According to the discussion in Section \ref{Dirac}, 
there remains 5 textures which should be examined. 
However, due to the 1-2 permutation invariance of $M_C^{-1}$,
we need not try the 3 textures which are related with (\ref{2zeromDb})
by exchanging the columns.
Thus it is enough to investigate the two textures which
are obtained by exchanging the 1-3 and 2-3 rows of (\ref{2zeromDb}),
as in the case for the normal hierarchy.
For the Dirac mass matrix $m_D$ which is obtained by 1-3 row exchange
of (\ref{2zeromDb}), the relation (\ref{a38}) implies that the small
entry of $\mathcal{O}(10^{-3})$ is sitting in $V_{33}$.
Moreover, the element $|V_{23}|$ must be near maximal to fit the atmospheric
neutrino data.
Thus the 1-3 element $|V_{13}|$ is also near maximal, which is by no means
viable.
On the other hand the 2-3 exchanging texture is apparently viable
because the bi-maximal mixing does not change physics 
under this exchange.

\subsection{Zeros of $m_D$ and $M_R$ combined --  viable mass matrices --} 
In this section, we explore the case where the neutrino Yukawa sector
has total 3 vanishing elements.
In this level, we will find textures which are totally compatible
with the data and have one definite correlation among neutrino masses
and mixings.

For total 3 zero case the different possibilities are 

\begin{enumerate} 
\item[(A)] 
3 zero $m_D$ and no zero $M_R$ 

\item[(B)] 
2 zero $m_D$ and one zero $M_R$

\item[(C)] 1 zero $m_D$ and two zero $M_R$

\end{enumerate} 
By exhausting all texture combinations in each category above,
we found 7 combinations of
$m_D$ and $M_R$ which are consistent with the present $3\sigma$
data for the three generation neutrino oscillation.

Table \ref{3zeroR}. shows 7 solutions and their predictions.
Besides the 7 patterns in the table, there exist other 7 solutions
which can be obtained by permuting 2-3 rows of $m_D$ for each texture.
Although these 14 patterns are independent in the sense that they are
not related each other by the field rotation, the predictions are
almost the same for both 7 textures.
Thus we present only 7 partners in Table \ref{3zeroR}.
We would like to emphasize that we are not dropping any possibilities.
The result is obtained partly by general considerations and
partly by direct examination of each mass matrix combination.

In the following, we build up the whole picture by discussing each of the 
seven cases presented in Table \ref{3zeroR}.

\begin{table}
\begin{center}
\begin{tabular}{c|c|c|c|c|c}\hline\hline
& $m_D$, ~~~~$M_R^{-1}$ & NH & IH & $\sin\theta_{13}$ &
$ |m_{ee}| $ (eV)
 \\\hline
A1
&
$
\begin{pmatrix}
0 & d \\ b & 0 \\ c & 0 \\
\end{pmatrix}$,
$\begin{pmatrix}
A & B \\ B & C \\
\end{pmatrix} $
& $\times$ & $\bigcirc$ & $\sim 0$ & $\sim$ 0.02
\\\hline\hline
B1
&
$
\begin{pmatrix}
a & d \\ b & 0 \\ c & 0 \\
\end{pmatrix}$,
$\begin{pmatrix}
A & 0 \\ 0 & C \\
\end{pmatrix} $
& $\times$ & $\bigcirc$ & $\sim 0$ & $\sim$ 0.02
\\\hline
B2
&
$
\begin{pmatrix}
a & 0 \\ b & e \\ 0 & f \\
\end{pmatrix}$,
$\begin{pmatrix}
A & 0 \\ 0 & C \\
\end{pmatrix} $
& $\bigcirc$ & $\bigcirc$ &
$
\left\{ \begin{array}{ll}
\simeq \frac{1}{2}\sin2\theta_{12}\tan\theta_{23}
\sqrt{\alpha},& {\rm (NH)} \\
\simeq \frac{1}{4}\sin2\theta_{12}\tan\theta_{23}\,\alpha, & {\rm (IH)} \\
\end{array} \right.
$
&
$
\left\{ \begin{array}{ll}
\sim 0.003 ,& {\rm (NH)} \\
\sim 0.05,& {\rm (IH)} \\
\end{array} \right.
$
\\\hline
B3
&
$
\begin{pmatrix}
a & d \\ b & 0 \\ c & 0 \\
\end{pmatrix}$,
$\begin{pmatrix}
A & B \\ B & 0 \\
\end{pmatrix} $
& $\times$ & $\bigcirc$ & $\sim 0$ & $\sim$ 0.02
\\\hline
B4
&
$
\begin{pmatrix}
a & 0 \\ 0 & e \\ c & f \\
\end{pmatrix}$,
$\begin{pmatrix}
0 & B \\ B & C \\
\end{pmatrix} $
& $\bigcirc$ & $\times$ &
$\simeq  \alpha^{\frac{1}{4}}
\sin\theta_{12}
$
& $\sim$ 0
\\\hline\hline
C1
&
$
\begin{pmatrix}
a & 0 \\ b & e \\ c & f \\
\end{pmatrix}$,
$\begin{pmatrix}
0 & B \\ B & 0 \\
\end{pmatrix} $
& $\bigcirc$ & $\times$ &
$\simeq \alpha^{\frac{1}{4}}
\sin\theta_{12}
$
& $\sim$ 0
\\\hline
C2
&
$
\begin{pmatrix}
a & d \\ b & 0 \\ c & f \\
\end{pmatrix}$,
$\begin{pmatrix}
                                         0 & B \\ B & 0 \\
\end{pmatrix} $
& $\times$ & $\bigcirc$ &
$\simeq \left( 1 - \sqrt{2}\sin\theta_{12} - \frac{1}{8}\alpha \right)
\cot\theta_{23}
$
& $\sim$ 0.01
\\\hline\hline
\end{tabular}
\end{center}
\caption{The 7 solutions for the total 3 zero textures.
The column ``NH'' and ``IH'' means the normal and the inverted
hierarchy respectively.
In these columns, the symbol ``$\bigcirc$'' means each texture can accommodate
each mass ordering, and ``$\times$'' means it cannot.
For the column ``$\sin\theta_{13}$'', we show the correlations 
between $\sin\theta_{13}$ and other observables
to the leading order of $\alpha \equiv \Delta m_{21}^2/|\Delta m_{31}^2|$ 
and $\sin\theta_{13}$.
For the column $|m_{ee}|$, we show a typical magnitudes 
for the averaged neutrino masses responsible for neutrino-less double
beta decay.}
\label{3zeroR}
\end{table}

\subsubsection{A1. 3 zero $m_D$ and 0 zero $M_R$}

Here we discuss the solution A1, where the Dirac mass matrix $m_D$ has
3 zeros while the Majorana mass matrix $M_R$ has no-vanishing
entries;
\begin{eqnarray}
m_D = \begin{pmatrix}
0 & d \\ b & 0 \\ c & 0 \\
\end{pmatrix},\quad
M_R^{-1} = \begin{pmatrix}
A & B \\ B & C \\
\end{pmatrix}.
\label{3md0mc}
\end{eqnarray}
For this case we have 12 real parameters characterizing $m_D$ 
and $M_R$. All the 3 phases of $m_D$ and 2 phases of $M_R$ 
can be removed by redefining 
the fields and so for this case we have 7 parameters -- 3 real 
parameters for $m_D$ and 4 real parameters for $M_R$ (one of which 
is a phase).  
As we discussed below (\ref{aa}), we can redefine the lepton fields
in such a way that only one parameter in $M_R$ has complex phase.
Following the discussion below (\ref{aa}), here also we put
the phase to the matrix element $C$.

We note that this case has already been discussed 
as a part of the general discussion for 2 zero $m_D$ and 1 zero $M_R$ 
(cf. eq. (\ref{aa})) and the conclusions obtained {\it i.e.} 
$\theta_{13} =0$, inverted hierarchy and $\theta_{23} = \pi/4$ 
for $b=c$ are all applicable here. 
After this general discussion we proceeded with detailed calculation 
for the $C = 0$ case and found that the 4 zero texture cannot 
give the correct solar mixing. 
In this section we study the effect of $C \neq 0$. 
In order to study the effect of the $d^2 C$ term, let us change
the coordinate by orthogonal transformation $\widetilde{U}$;
\begin{eqnarray}
\widetilde{U} = \begin{pmatrix}
1 & 0 & 0 \\
0 & \frac{b}{\sqrt{b^2 + c^2 }} & \frac{-c}{\sqrt{b^2 + c^2 }} \\
0 & \frac{c}{\sqrt{b^2 + c^2 }} & \frac{b}{\sqrt{b^2 + c^2 }} \\
\end{pmatrix}.
\label{u23a1}
\end{eqnarray}
Then we find
\begin{eqnarray}
\widetilde{\mathcal{M}}
&=& \widetilde{U}^{\rm T}\mathcal{M} \widetilde{U} \nonumber\\
&=&\begin{pmatrix}
d^2 C & \sqrt{b^2 + c^2}dB & 0 \\
\sqrt{b^2 + c^2}dB & (b^2 + c^2)A & 0 \\
0 & 0 & 0 \\
\end{pmatrix}.
\label{mtilde}
\end{eqnarray}
Note that we are choosing the basis in which only the matrix element
$C$ has a complex phase.
Thus we can regard only $\mathcal{M}_{11}$ as complex valued without
loss of generality.
In this basis, the effect of $C \neq 0$ is clear.
If the element $C$ were vanishing, there would be correlation
between the mass eigenvalues and the 1-2 mixing angle
in the upper-left $2\times 2$ matrix of (\ref{mtilde}).
However in this case the existence of $d^2C$ term relaxes the constraint
and we can fit any mass eigenvalues and the 1-2 mixing by tuning
the parameters in (\ref{mtilde}).

While there is one complex phase in (\ref{3md0mc}) and (\ref{mtilde}),
we have no chance to observe CP violation in the oscillation experiments
as $\theta_{13} = 0$. 
However leptogenesis \cite{leptogenesis1} is possible by this phase.
In the limit $\theta_{13} = 0$ the effective mass constrained by 
$0\nu 2\beta$  decay ($|d^2 C|$) can be expressed as  
\be
|m_{ee}| \simeq \sqrt{|\Delta m^2_{31}|} 
\sqrt{ 1 - \sin^2 2\theta_{12} \sin^2\left( \rho/2 \right)}
\label{ihth130}
\ee 
which gives 
\, $\approx$ \,  $0.02 - 0.05$ eV,
where the two limits correspond to  $\rho = \pi$ and $\rho = 0$ respectively.  
Thus 
$0\nu 2\beta$ will be observed in future
if this texture is realized in nature \cite{future2beta}. 

Another interesting feature of (\ref{3md0mc}) is that it provides
tri-bimaximal mixing \cite{tribi} under the condition where $b =c$ and 
$d^2 C = 2b^2A - bd B$ hold.
Although the latter condition implies a 
nontrivial correlation between Dirac and Majorana mass matrix, 
to build models which realize these relations might be an interesting
research direction.

We note that this texture, which can be consistent with 
low energy phenomenology, is obtained by the choice of a non-diagonal 
$M_R$. In the $M_R$ diagonal basis the following relationships will 
hold between the elements of $m_D$: 
$-m_{D11}/m_{D12} = m_{D22}/m_{D21} = m_{D32}/m_{D31}= \tan \theta_R$, where, 
$\theta_R$ is the angle that parametrizes the matrix $U_R$ in  
(\ref{mrdiagbasis}) that diagonalizes $M_R$ and can in general
be expressed as
\begin{eqnarray}
U_R = \begin{pmatrix}
\cos \theta_R & \sin \theta_R \\ -\sin\theta_R & \cos\theta_R  \\
\end{pmatrix} ,\quad
\label{ur}
\end{eqnarray}
Hence if one considers texture zeros in $m_D$ with a diagonal $M_R$  
this texture will not get included.

\subsubsection{B1. Two zero $m_D$ and 1 zero $M_R$}

Here we discuss the solution B1, where the Dirac mass matrix $m_D$ has
2 zero while the Majorana mass matrix $M_R$ has 1 zero
entry;
\begin{eqnarray}
m_D = \begin{pmatrix}
a & d \\ b & 0 \\ c & 0 \\
\end{pmatrix} ,\quad
M_R^{-1} = \begin{pmatrix}
A & 0 \\ 0 & C \\
\end{pmatrix}.
\label{2.22}
\end{eqnarray}
After the seesaw integration, we find
\begin{eqnarray}
\mathcal{M} =
\begin{pmatrix}
a^2 & ab & ac \\
ab & b^2 & bc \\
ac & bc & c^2 \\
\end{pmatrix}A
+
\begin{pmatrix}
d^2 & 0 & 0 \\
0 & 0 & 0 \\
0 & 0 & 0 \\
\end{pmatrix}C.
\label{2.23}
\end{eqnarray}
Note that we can take a basis in which only the parameter $a$ or $d$
have complex phase with no loss of generality.
In spite of the existence of this un-removable phase, there is
no CP violation at the low energy.
One can understand this by noting that this matrix has an eigenvector
$(0,\,-c/\sqrt{b^2 + c^2}\, ,\, b/\sqrt{b^2 + c^2}\,)^{\rm T}$.
Thus the same discussion holds as in the case of the solution A1
and the low energy predictions are similar.

However, the texture A1 and B1 are indeed independent.
They are not associated with each other by the unitary transformation
of the fields.
The physical difference can arise at some high energy scale where
the right-handed neutrinos are active.
For example, leptogenesis or renormalization group effect-induced
lepton flavor violation in supersymmetry 
can have different implications in the two scenarios. 

It is worth mentioning that the tri-bimaximal mixing is realized
if $c = b$ and $a^2A + d^2 C = 2 b^2 A - abA$ hold.
By trying to build models which realize this relation,
we might gain insights into underlying symmetry or
dynamical mechanism for the generation structure.

\subsubsection{B2. Two zero $m_D$ and one zero $M_R$}

We now consider the case where the 2 zeros in $m_D$ are in different columns
and the one zero in the Majorana Mass matrix is in the off-diagonal position 
{\it i.e.} the Majorana mass matrix $M_R$ is diagonal;
\begin{eqnarray}
m_D = \begin{pmatrix}
a & 0 \\ b & e \\ 0 & f \\
\end{pmatrix} ,\quad
M_C^{-1} = \begin{pmatrix}
A & 0 \\ 0 & C \\
\end{pmatrix}.
\label{2.24}
\end{eqnarray}
This texture has been extensively discussed in literatures from
various point of view \cite{2nuR}.
Since the Majorana mass matrix is diagonal, 
only the Dirac mass matrix $m_D$ is responsible for the generation
mixing. 
After the seesaw integration, we find
\begin{eqnarray}
\mathcal{M} =
\begin{pmatrix}
a'^2 & a'b' & 0 \\
a'b' & b'^2+ c'^2 & c'd' \\
0 & c'd' & d'^2 \\
\end{pmatrix},
\label{2.25}
\end{eqnarray}
where we re-defined the parameters as $a' \equiv a\sqrt{A}$,
$b' \equiv b\sqrt{A}$, $c' \equiv e\sqrt{C}$ and $d' \equiv f\sqrt{C}$.
For the above case there are in general 12 parameters (6 angles and 6 phases). 
5 phases can be absorbed in the neutrino fields. Therefore one eventually has 
12 parameters (6 real parameters and 1 phase) . 
Note that we can take a basis in which only the parameter $b$ or $e$
have complex phase with no loss of generality.

One of the most striking feature of this texture is that it can accommodate
the normal and the inverted hierarchy simultaneously.
In Table \ref{3zeroR}, this is the only solution which has such a
strong flexibility.
In the following we shall discuss each case in detail individually.

\paragraph{Normal hierarchy}

In the Majorana mass matrix (\ref{2.25}), the large mixture for the
atmospheric data and the normal mass ordering can be naturally accommodated
by taking $c' \simeq d' $ and $c'd' \gg a'b'$.
The solar angle can be nicely fitted by tuning $a'$ and $b'$.
One of the most interesting features for this texture is that there is
a connection between CP violation phenomena at high energy and low energy.
Also interesting is that the prediction for $\theta_{13}$ which
originates in the vanishing elements in 1-3 position;
\begin{eqnarray}
\sin\theta_{13} &\,\simeq\,& \frac{1}{2}\sin2\theta_{12}\tan\theta_{23}
\sqrt{\alpha } \nonumber\\
&\, = \,& 0.050 - 0.14,
\label{13predic}
\end{eqnarray}
where $\alpha \equiv \Delta m_{21}^2/\Delta m_{31}^2$.
In the last line, we substitute $3\sigma$ boundary values
into each observables.
The predicted range is very encouraging for the next generation
oscillation experiments with artificial sources \cite{dc}. 

The effective mass $|m_{ee}|$ governing neutrino-less double beta decay 
for normal hierarchy can be expressed as 
\be
|m_{ee}| \,=\, \left| e^{i(2 \delta + \sigma - \rho)} 
\sqrt{\Delta m^2_{21}} c_{13}^2 s_{12}^2 + \sqrt{\Delta m^2_{31}} s_{13}^2
\right|. 
\label{meenh}
\ee 
From the relation (\ref{13predic}), we can see that the contribution 
from the first term is dominant in (\ref{meenh}).
By substituting best fit values of the solar angle and the mass differences,
we obtain $|m_{ee}| \sim 0.003$ eV for the normal hierarchy. 

\paragraph{Inverted hierarchy}

On the other hand, (\ref{2.25}) can also accommodate the inverted
mass ordering 
by taking
\begin{eqnarray}
&&a'^2 \,\simeq\, \sqrt{|\Delta m_{31}^2|},\\
&&a'b' \,\simeq\, \sqrt{|\Delta m_{31}^2| - \Delta m_{21}^2} -
\sqrt{|\Delta m_{31}^2|},\\
&&b'^2 + c'^2  \,=\, c'd'
\,=\, d'^2
\,\simeq\, \frac{\sqrt{|\Delta m_{31}^2|}}{2}.
\end{eqnarray}
After diagonalizing the 2-3 block of (\ref{2.25}) by maximal mixing,
we find the small 1-3 (and 3-1) element appears as $\sim a'b'/\sqrt{2}$.
Thus we can naively estimate the prediction for the reactor angle
to be
\begin{eqnarray}
\sin\theta_{13}
 &\simeq& \frac{1}{2\sqrt{2}}
\frac{\Delta m_{21}^2}{|\Delta m_{31}^2|}
\,\sim\, 0.01,
\label{2.31}
\end{eqnarray}
where we use the best fit values for the mass differences
in the last line.
Unfortunately the predicted magnitude is small compared to the
possible reach of the next generation experiments.

It is interesting to notice that the sum of the matrix
elements in each row in (\ref{2.25}) is nearly the same.
That is, $a'^2 + a'b' \simeq a'b' + b'^2 + c'^2 + c'd' \simeq
c'd' + d'^2 \simeq \sqrt{|\Delta m_{31}^2|}$.
This implies that we can have a tri-maximal eigenvector
$(1/\sqrt{3} \,,\,1/\sqrt{3}\,,\,1/\sqrt{3})^{\rm T}$ by tuning the
parameters.
Thus, the mixing matrix must be made out of the product of the tri-bimaximal
mixing and a perturbation matrix;
\begin{eqnarray}
V &=& V_{\rm tri} \,O_\epsilon \nonumber\\
&=&
\begin{pmatrix}
\frac{-2}{\sqrt{6}} & \frac{1}{\sqrt{3}} & 0 \\
\frac{1}{\sqrt{6}} &  \frac{1}{\sqrt{3}} & \frac{-1}{\sqrt{2}} \\
\frac{1}{\sqrt{6}} &  \frac{1}{\sqrt{3}} & \frac{1}{\sqrt{2}} \\
\end{pmatrix}
\begin{pmatrix}
\cos\theta_\epsilon & 0 & \sin\theta_\epsilon \\
0 & 1 & 0 \\
-\sin\theta_\epsilon & 0 & \sin\theta_\epsilon \\
\end{pmatrix},
\label{2.32}
\end{eqnarray}
where the angle $\theta_\epsilon$ is small as we will see in what follow.
From this expression, we can infer that some generation symmetry exists
behind the mass matrix (\ref{2.25}).
For example, it is known that the $S_3$ flavor symmetry and its breakdown can
naturally produce the mixing form of (\ref{2.32}) \cite{s3}.
If such a symmetric structure can account for the texture (\ref{2.25}),
the parameter set which is needed to produce the inverted hierarchy
is no longer a group of tuned-parameters but rather unavoidable consequence
of flavor symmetry and its breaking phenomena.
Furthermore, we can use the expression (\ref{2.32}) for more
practical purposes.
For example, we can obtain more precise expression
for $\theta_{13}$ than the rough estimation (\ref{2.31})
by fixing the angle $\theta_\epsilon$ in (\ref{2.32}).
By reconstructing the mass matrix in terms of the $V$ in
(\ref{2.32}) and the diagonal mass eigenvalues
${\rm diag}(\sqrt{|\Delta m_{31}^2|}\, , \,
\sqrt{|\Delta m_{31}^2| + \Delta m_{21}^2} \, , 0)$, and
imposing the texture zero condition for the 1-3 entry,
we find
\begin{eqnarray}
\frac{m_2}{3} \,- \,
\sqrt{\frac{2}{3}}\cos\theta_\epsilon \left(
\frac{\cos\theta_\epsilon}{\sqrt{6}} -
\frac{\sin\theta_\epsilon}{\sqrt{2}} \right)m_1
\, = \, 0,
\label{2.33}
\end{eqnarray}
where $m_1 = \sqrt{|\Delta m_{31}^2|}$ and
$m_2 = \sqrt{|\Delta m_{31}^2| + \Delta m_{21}^2}$.
From this equation we can fix the perturbation $\theta_\epsilon$ as
\begin{eqnarray}
\sin\theta_{\epsilon}
&\, \simeq \, & \frac{\alpha}{2\sqrt{3}},
\label{2.34}
\end{eqnarray}
where $\alpha \equiv \Delta m_{21}^2/|\Delta m_{31}^2|$.
Here we choose the smaller solution of the equation (\ref{2.33})
for fitting the $\theta_{13}$.
From (\ref{2.34}) we find that $\sin\theta_{13} = \frac{\alpha}{3\sqrt{2}}$
which agrees well with (\ref{2.31}).

The above prediction is interesting in the sense that it is associated
with the tri-bimaximal mixing and its deviation.
However we can expand (\ref{2.34}) to more general formula.
As we have done in the case of the normal hierarchy, we can
write the Majorana mass matrix in terms of the mixing angles and
the mass eigenvalues, and derive the condition for the 1-3 and 3-1
vanishing elements.
Then we find
\begin{eqnarray}
-m_1 c_{12}^2 s_{13} \cos\delta + m_1 c_{12}s_{12}t_{23}
- m_2 s_{12}^2 s_{13}\cos( \delta + \rho )
- m_2 c_{12}s_{12} t_{23} \cos\rho = 0,
\label{re2}
\end{eqnarray}
and
\begin{eqnarray}
-m_1 c_{12}^2 s_{13} \sin\delta 
- m_2 s_{12}^2 s_{13}\sin( \delta + \rho )
- m_2 c_{12}s_{12} t_{23} \sin\rho = 0.
\label{im2}
\end{eqnarray}
where $t_{23} \equiv \tan\theta_{23}$, $\delta$ is the Dirac phase and
$\rho$ stands for the Majorana phase.
From (\ref{im2}), we can see that $\delta = \rho = \mathcal{O}(s_{13})$ must hold
in order to keep the texture zero in 1-3 position.
Then (\ref{re2}) gives
\begin{eqnarray}
\sin\theta_{13} \,=\, \frac{\alpha}{4} \sin 2\theta_{12} \tan\theta_{23}
\end{eqnarray}
as a leading order relation in $\alpha$ and $\sin\theta_{13}$.
Note that at the tri-bimaximal limit 
$\sin 2\theta_{12} \to \frac{2\sqrt{2}}{3} $ and 
$\tan\theta_{23} \to 1$, we can reproduce the previous formula
of $\sin\theta_{13} = \frac{\alpha}{3\sqrt{2}}$.
For inverted hierarchy, 
the effective mass $m_{ee}$ measured in neutrino-less double beta decay 
can be expressed as in (\ref{ihth130}),
which gives relatively large mass parameter $|m_{ee}| \sim 0.05$ eV 
because of the small Majorana phase of $\rho = \mathcal{O}(s_{13})$.

The CP violating phases and their connection to baryon number asymmetry 
of the universe for this texture was considered in literature
in \cite{2nuR}.

\subsubsection{B3. Two zero $m_D$ and one zero $M_R$}

Here we discuss the solution B3, where the Dirac mass matrix $m_D$ has
2 zero while the Majorana mass matrix $M_C$ has 1 zero
entry;
\begin{eqnarray}
m_D = \begin{pmatrix}
a & d \\ b & 0 \\ c & 0 \\
\end{pmatrix} ,\quad
M_C^{-1} = \begin{pmatrix}
A & B \\ B & 0 \\
\end{pmatrix}.
\label{2.35}
\end{eqnarray}
After the seesaw operation, we find
\begin{eqnarray}
\mathcal{M} =
\begin{pmatrix}
a^2 & ab & ac \\
ab & b^2 & bc \\
ac & bc & c^2 \\
\end{pmatrix}A
+
\begin{pmatrix}
2ad & bd & cd \\
bd & 0 & 0 \\
cd & 0 & 0 \\
\end{pmatrix}B.
\label{2.36}
\end{eqnarray}
Note that we can take a basis in which only the parameter $a$ or $d$
have complex phase with no loss of generality.
It is useful to notice that this matrix has an eigenvector
$(0,\,-c/\sqrt{b^2 + c^2}\, ,\, b/\sqrt{b^2 + c^2}\,)^{\rm T}$.
Thus the same discussion, as in the case of the solution A1
and B1, remain valid leading to similar low energy predictions.

As the solutions A1 and B1, this texture can also provide
the tri-bimaximal mixing. 
It is realized if $c = b$ and $a^2A + 2adB = 2 b^2 A - (abA + bdB)$ hold.

Note that in this case the $M_R$ diagonal basis implies 
the following relation between the elements 
$m_{D22}/m_{D21} = m_{D32}/m_{D31} = \tan \theta_R$  where 
$\theta_R$ is the angle parametrizing  
$U_R$  as in (\ref{ur}).  
Thus in the $M_R$ diagonal 
basis the zeros of $m_D$ will not be visible. 

\subsubsection{B4. Two zero $m_D$ and one zero $M_R$ }

Here we discuss the texture B4, where
the Majorana mass matrix has texture zero in 1-1 position;
\begin{eqnarray}
m_D = \begin{pmatrix}
a & 0 \\ 0 & e \\ c & f \\
\end{pmatrix} ,\quad
M_C^{-1} = \begin{pmatrix}
0 & B \\ B & C \\
\end{pmatrix}.
\label{2.37}
\end{eqnarray}
For this case there are 8+4 = 12 real parameters characterizing $m_D$ 
and $M_R$. The 2 phases in $M_R$ 
can be absorbed by redefining the fields and one can consider $M_R$ 
to be real. 3 phases in $m_D$ can likewise be removed leaving 
5 parameters one of which is a phase. 
Note that we can take a basis in which only the parameter $c$ or $f$
have complex phase with no loss of generality.
After the seesaw integration, we find the induced Majorana mass
matrix to be
\begin{eqnarray}
\mathcal{M} =
\begin{pmatrix}
0 & a'e' & a'f' \\
a'e' & 0 & c'e' \\
a'f' & c'e' & 2c'f' \\
\end{pmatrix} +
\begin{pmatrix}
0 & 0 & 0 \\
0 &  e'^2 & e'f' \\
0 & e'f' & f'^2 \\
\end{pmatrix}
\label{2.38},
\end{eqnarray}
where the right-handed parameters are absorbed as
$e' \equiv e\sqrt{C}$, $f' \equiv f\sqrt{C}$,
$a' \equiv aB/\sqrt{C}$, $c' \equiv cB/\sqrt{C}$.

From (\ref{2.38}), we can see that 4 effective parameters
$a', c',  e', f'$ control the low-energy physics.
Thus there should be one relation among the 5 observables.
As a consequence of the fact that the 1-1 element is vanishing
and $m_1 = 0$, we have the prediction 
\begin{eqnarray}
 \sin\theta_{13} \,\simeq\,
\alpha^{\frac{1}{4}}
 \sin\theta_{12}
\label{2.39},
\end{eqnarray}
which is the same as the prediction of the 4 zero textures (\ref{4zeroB}).
For the texture (\ref{4zeroB}), we have found that $\sin\theta_{23}$
is predicted to be too small compared to the observed large mixture.
However, in the texture (\ref{2.38}), 
we have a correction matrix in 2-3
sector against the first term which is the same form as (\ref{4zeroB}).
We already have one relation of (\ref{2.39}) so that $\sin\theta_{23}$
can be fitted by tuning the original mass matrix parameters.
The $|m_{ee}|$ is $\sim$ 0 in this case 
which is beyond the reach of the 
next generation neutrino-less double beta experiments. 

As we have discussed in the texture (\ref{4zeroB}), the correlation
(\ref{2.39}) requires marginal values of the mass differences
and the solar angle in order that the reactor angle is inside 
the present $3\sigma$ allowed range.
Together with the improvement of the data about $\theta_{13}$,
precise measurements for the mass differences and the solar angle
can judge this texture in near future. 

In this case the $M_R$ diagonal basis would imply 
the following relation between the elements 
$-m_{D21}/m_{D22} = m_{D12}/m_{D11} =  \tan \theta_R$  where 
$\theta_R$ is the angle parametrizing  
$U_R$  as in (\ref{ur}).  

\subsubsection{C1. One zero $m_D$ and two zero $M_R$}

Here we discuss the texture C1, where
the Dirac mass matrix has 1 zero and
the Majorana mass matrix has 2 zeros;
\begin{eqnarray}
m_D = \begin{pmatrix}
a & 0 \\ b & e \\ c & f \\
\end{pmatrix} ,\quad
M_C^{-1} = \begin{pmatrix}
0 & B \\ B & 0 \\
\end{pmatrix}.
\label{2.40}
\end{eqnarray}
In this case there are  6  parameters (1 of which is phase) 
characterizing $m_D$ and 1 parameter for $M_R$. 
We can take a basis in which only $b$, $e$, $c$ or $f$ has
complex phase with no loss of generality.
After the seesaw integration, we find
\begin{eqnarray}
\mathcal{M} =
\begin{pmatrix}
0 & a'e' & a'f' \\
a'e' & 0 & c'e' \\
a'f' & c'e' & 2c'f' \\
\end{pmatrix} +
\begin{pmatrix}
0 & 0 & 0 \\
0 &  2e'b' & b'f' \\
0 & b'f' & 0 \\
\end{pmatrix}
\label{2.41},
\end{eqnarray}
where the right-handed parameters are absorbed as
$a' \equiv a\sqrt{B}$, $b' \equiv b \sqrt{B}$, $c' = c\sqrt{B}$,
$e' \equiv e\sqrt{B}$

It is immediately seen that the texture zero in the 1-1 position
leads to the same prediction as that of B4 for the normal hierarchy
(\ref{2.39}).
Moreover, as in the texture B4, we have a correction matrix
in 2-3 sector against the Majorana mass matrix which is the resultant
form of (\ref{4zeroB}) after seesaw operation, though the structure
of the correction is different from that of B4.
It turns out however that the second term (\ref{2.41}) can also
help to fit $\sin\theta_{23}$, so that there is no difference
between the low energy predictions of B4 and C1.

In this case the $M_R$ diagonal basis would imply 
the following relation between the elements of $m_D$:
$m_{D12}/m_{D11} =  \tan \theta_R$  where 
$\theta_R$ is the angle parametrizing  
$U_R$  as in (\ref{ur}).  

\subsubsection{C2: One zero $m_D$ and two zero $M_R$}

Here we discuss the texture C2, which is the last option in Table \ref{3zeroR};
\begin{eqnarray}
m_D = \begin{pmatrix}
a & d \\ b & 0 \\ c & f \\
\end{pmatrix} ,\quad
M_C^{-1} = \begin{pmatrix}
0 & B \\ B & 0 \\
\end{pmatrix}.
\label{C2tex}
\end{eqnarray}
This texture is obtained by exchanging 1-2 rows of $m_D$ in (\ref{2.40}),
so that we can obtain the low-energy Majorana mass matrix by
permuting the 1-2 rows and the columns of (\ref{2.41}) with replacements
$b \to a$, $a \to b$ and $e \to d$.
As in the solution C1, we can take the basis in which the matrix element
$a$, $d$, $c$ or $f$ has complex phase.

It turns out that this texture is viable only with the inverted hierarchy.
Since the effective Majorana mass matrix is obtained by 1-2 exchange
of the rows and the columns of (\ref{2.41}), there is one texture zero
in 2-2 position.
We thus have one correlation among masses and mixings.
We find the sum rule
\begin{eqnarray}
\sin\theta_{12} \, \simeq \, \frac{1}{\sqrt{2}} \,-\, 
\frac{1}{8\sqrt{2}}\alpha
\,\,- \,\, \frac{1}{\sqrt{2}}\tan\theta_{23}\sin\theta_{13}
\label{sumrule}
\end{eqnarray}
holds.
Here $\alpha \equiv \Delta m_{21}^2/|\Delta m_{31}^2|$
and this relation is given as the leading order approximation
in powers of $\alpha$ and $\sin\theta_{13}$.
The phase parameters are to be fixed as $\rho = \pi$ and
$\delta = 0$ up to this order.
Note that we can see this equation as a modified formula
for (\ref{v12pre}) which leads to too large solar mixing.
Now the problem is ameliorated by the existence of the third
term.
If $\theta_{13}=0$ then this texture cannot 
give the correct solar angle. 

By setting $\alpha \to 0$ and $\tan\theta_{23} \to 1$,
we find (\ref{sumrule}) implies
\begin{eqnarray}
\sin\theta_{13} \,\simeq\, 1 - \sqrt{2}\sin\theta_{12} \,=\, 0.20^{+ 0.08}_{-0.1}.
\end{eqnarray}
In the right hand side we used the best fit and $3\sigma$ boundary values
for the $\sin\theta_{12}$.
The reactor angle must be just below the present $3\sigma$ upper bound.
The future reactor experiment will confirm $\sin\theta_{13}$ of 
$\mathcal{O}(10^{-1})$ if this texture is realized.

Moreover, it is interesting to note that 1-1 element of $\mathcal{M}$
is relatively large because of the inverted mass spectrum:
\begin{eqnarray}
|m_{ee}| \, \simeq \, 
\sqrt{|\Delta m_{31}^2|}\cos 2\theta_{12} \,\sim\, 0.02 \,\,{\rm eV}.
\end{eqnarray}
Thus, with this texture form, we have a good chance to confirm 
that the neutrinos are indeed Majorana particles in near future 
\cite{future2beta}.
The most prominent feature is that in both measurements -- the reactor
angle and neutrino-less double beta decay -- we will find positive
signals simultaneously at the next generation facilities.

Finally we would like to comment on the CP violation.
As we mentioned above, there is one phase which cannot be removed
by the redefinition of the fields.
This one phase controls all CP violation phenomena at high energy
as well as low energy.
It is interesting to observe that the heavy neutrino masses
are degenerate in (\ref{C2tex}).
Although the lepton asymmetry vanish with exact degeneracy,
the degeneracy may be relaxed, for example, by radiative corrections
from the other sector.
This fact may lead to enhanced lepton asymmetry by the contribution
from the self-energy diagram \cite{leptogenesis2}.
The baryon number of the universe will be proportional to
${\rm Im} (m_D^\dag m_D)_{12}^2 = -2ad\mathcal{X}\sin\phi$, where
$\phi$ is the phase (of $a$ or $d$) and 
$\mathcal{X} = |a|^2 + |b|^2 + |c|^2 -|d|^2 - |f|^2$.
On the other hand the weak basis invariant \cite{lowCP} responsible for 
low energy CP violation is found to be proportional to
$(\mathcal{F}\cos\phi + \mathcal{G})\sin\phi$ where 
$\mathcal{F}$ and $\mathcal{G}$ are some functions of the elements
of $m_D$.
The detailed analysis of CP violation and the models which produce
the texture (\ref{C2tex}) will be presented in a separate paper 
\cite{future}.

In this case the $M_R$ diagonal basis would imply 
the following relation between the elements of $m_D$:
$m_{D22}/m_{D21} =  \tan \theta_R$  where 
$\theta_R$ is the angle parametrizing  
$U_R$  as in (\ref{ur}).  

\section{Summary and Conclusions} 

In this paper we analyze the texture zeros in the 
neutrino Yukawa Coupling matrix $m_D$ and the heavy Majorana neutrino 
mass matrix $M_R$ in the context of the minimal seesaw model 
including 2 heavy right-handed neutrinos. 
We illustrate which textures are compatible with the present neutrino 
oscillation data and discuss their implications for the future 
neutrino experiments.
We do not make the assumption that $M_R$ is diagonal. 

We first consider the zeros in the neutrino Dirac mass matrix $m_D$ and 
show that it cannot have 4 or more zeros. 
Thus the maximum number of zeros $m_D$ can have is 3. 
For the Majorana mass matrix of the right-handed neutrinos $M_R$, 
the maximum number of zeros can be 2.
Except for the case where $M_R$ has vanishing determinant,
the possible texture in this case is zeros in the diagonal position. 
Thus the maximal number of zeros that can be allowed in $m_D$ and 
$M_R$ taken together are 5 (3 zeros in $m_D$ and 2 zeros in $M_R$). 
But such a pattern give rise to a Majorana mass matrix with 
more than two zeros which is incompatible with data according to 
\cite{Frampton:2002yf}.

If we consider total 4 zeros in 
$m_D$ and $M_R$  then 
there are two possibilities : 
\\
(i) 3 zeros in $m_D$ and 1 zero in $M_R$ \\
(ii) 2 zeros in $m_D$ and 2 zeros in $M_R$ \\
The pattern (i) gives rise to the inverted mass ordering with $m_3 = 0$, 
$\theta_{13} =0$ but the other angles are close to bimaximal mixing. 
The pattern (ii) can accommodate both normal and inverted hierarchy 
but it fails to predict one mixing angle.
For the normal hierarchy, the atmospheric angle $\theta_{23}$ is too large
or too small.
For the inverted hierarchy, it gives bi-maximal mixing which is not 
consistent with the current data which dictates $\theta_{12}$ to be close 
to $33^o$. 
Thus at this level also 
there are no acceptable solutions. 

At the next tier we consider total 3 zeros in $m_D$ and $M_R$. 
The possibilities in these cases are 
\\
(i) 3 zeros in $m_D$ and no zero in $M_R$ \\
(ii) 2 zeros in $m_D$ and 1 zero in $M_R$ \\
(iii) 1 zero in $m_D$ and 2 zeros in $M_R$ 
\\
By exhausting all texture possibilities in each category, 
we found seven patterns which are viable with the current data. 
This is summarized in Table \ref{3zeroR}. 
All these patterns can accommodate either normal hierarchy or  
inverted hierarchy or both and gives definite predictions 
for $\theta_{13}$ which can be testable in the near future. 

For case (i) with 3 zeros in $m_D$ and no zeros in $M_R$ it is possible 
to get inverted hierarchy with $\theta_{13} =0$. It is also possible to get 
tri-bimaximal mixing under certain conditions on the elements of 
$m_D$ and $M_R$.

For case (ii) there are two classes of patterns -- one in which both zeros 
in $m_D$ appear in the same column and the other in which the two zeros 
are placed  in different columns. 
In both cases assuming a diagonal $M_R$ it is possible to reproduce 
the current low energy data.  
The pattern with zeros in same column of $m_D$ can accommodate only
inverted hierarchy and predicts $\theta_{13} =0$. 
Again tri-bimaximal mixing is reproduced if certain equalities involving 
the elements of $m_D$ and $M_R$ are obeyed. 
The case where the zeros appear in different columns can accommodate 
both normal and inverted hierarchy. 
There are definite predictions for $\theta_{13}$ for both normal and inverted
case which can be tested in near future experiments.
In this category, we also include the possibility where one of the diagonal 
entries of $M_R$ is zero. 
In this situation, for the case where the two zeros of $m_D$ appear 
in the same column, one gets inverted hierarchy with $\sin\theta_{13}=0$. 
Whereas for $m_D$ with the 2 zeros in different columns one gets 
normal hierarchy with a definite non-zero prediction for 
$\theta_{13}$ in terms of $\Delta m^2_{21}/\Delta m^2_{31}$ and $\theta_{12}$. 

For case (iii) of 1 zero $m_D$ and 2 zero $M_R$  we find two 
allowed patterns viable with data. 
Both these cases are for non-diagonal $M_R$ with definite non-zero 
predictions for $\theta_{13}$. 

Summarizing, out of the 7 allowed patterns, 5 arise because we have 
relaxed the assumption of a diagonal $M_R$. 
In a basis where $M_R$ is diagonal the zeros in $m_D$ for all these 
5 patterns get hidden as specific relations between the different 
elements of $m_D$.  
To the best of our knowledge, these textures which are viable with the 
current data, have escaped attention since most analysis of 
texture zeros in literature had been done in a basis in which 
$M_R$ is diagonal, or avoided mentioning the mass textures which
do not have definite correlations among observable parameters. 

The mass matrix list of Table \ref{3zeroR} presents the most economical
extensions of the standard model.
The 7 solutions are minimal solutions for neutrino physics known at present
in terms of not only the number of the parameters but also the field
content. 
It is rather surprising that such simple economy solely leads to the rich
predictions shown in Table \ref{3zeroR}. 
Although the mass textures themselves can be discriminated only by
precise measurements of the low energy parameters,
the mechanism which realizes particular texture might provide
further predictions and/or new phenomena which can be targets
of next generation neutrino physics, collider physics, astrophysics and so on.
We hope that the 7 possibilities and preceded almost viable forms 
provide a foundation of model building which yields testable predictions 
and deeper understanding of the generation structure.

\bigskip\bigskip
\subsection*{Acknowledgments}

The authors acknowledge support from the neutrino project under the 
XI plan of the Harish-Chandra Research Institute. 
S.G. wishes to thank Anjan Joshipura for insightful discussions. 
A.W. is grateful to the organizers and participants of the conference
``NuHoRIzons'' held at Harish-Chandra Research Institute, Allahabad,
India on 13-15 February 2008, for discussions leading to this work.
The authors thank Probir Roy for his involvement in the initial 
phase of the work. 


\end{document}